# Deep-sequencing of the Peach Latent Mosaic Viroid Reveals New Aspects of Population Heterogeneity

Jean-Pierre Sehi Glouzon[1,2,a], François Bolduc[2,a], Rafael Najmanovich[2], Shengrui Wang[1], Jean-Pierre Perreault[2]*

[1]Département d'informatique, Faculté des sciences, Université de Sherbrooke, Sherbrooke, Québec, J1H 5N4, Canada.

[2]RNA Group/Groupe ARN, Département de biochimie, Faculté de médecine et des sciences de la santé, Pavillon de Recherche Appliquée au Cancer, Université de Sherbrooke, Sherbrooke, Québec, J1H 5N4, Canada.

[a]These authors contributed equally to this study.

**Running title: Genetic Variability of PLMVd**

Submitted: October 3rd, 2012

*Corresponding author:

Jean-Pierre Perreault, Ph.D (Jean-Pierre.Perreault@usherbrooke.ca)

Phone: (819) 564-5315; Fax: (819) 564-5340

# ABSTRACT

Viroids are small circular single-stranded infectious RNAs that are characterized by a relatively high mutation level. Knowledge of their sequence heterogeneity remains largely elusive, and, as yet, no strategy attempting to address this question from a population dynamics point of view is in place. In order to address these important questions, a GF305 indicator peach tree was infected with a single variant of the *Avsunviroidae* family member *Peach latent mosaic viroid* (PLMVd). Six months post-inoculation, full-length circular conformers of PLMVd were isolated, deep-sequenced and the resulting sequences analyzed using an original bioinformatics scheme specifically designed and developed in order to evaluate the richness of a given the sequence's population. Two distinct libraries were analyzed, and yielded 1125 and 1061 different PLMVd variants respectively, making this study the most productive to date (by more than an order of magnitude) in terms of the reporting of novel viroid sequences. Sequence variants exhibiting up to ~20% of mutations relative to the inoculated viroid were retrieved, clearly illustrating the high divergence dynamic inside a unique population. Using a novel hierarchical clustering algorithm, the different variants obtained were grouped into either 7 or 8 clusters depending on the library being analyzed. Most of the sequences contained, on average, between 4.6 and 6.3 mutations relative to the variant used initially to inoculate the plant. Interestingly, it was possible to reconstitute the sequence evolution between these clusters. On top of providing a reliable pipeline for the treatment of viroid deep-sequencing, this study sheds new light on the importance of the sequence variation that may take place in a viroid population and which may result in the formation of a quasi-species.

# AUTHOR SUMMARY

Viroids are small single-stranded circular RNA species that infect plants. They do not code for any protein, and are not protected by a protein capsid. They can escape plant cellular defenses and act as parasites, making them the smallest, self-replicating genetic element identified to date. Viroids have been shown to exhibit the highest mutation rate known, and it has been suggested that they evolve as a quasi-species since genomic heterogeneity is a characteristic found among all of the 30 known species. However, the extent of their sequence heterogeneity remains largely elusive, and strategies to uncover this knowledge from population dynamics have not yet been developed. In this study we explore the evolution of the *Peach latent mosaic viroid* following infection by a single sequence variant. Six months post-infection, the viroid's replicative conformer (i.e. the circular form) was isolated, cloned and then deep-sequenced in order to obtain a reliable picture of the heterogeneity of the related sequences present. In addition to providing a reliable pipeline for the treatment of viroid deep-sequencing, this study sheds new light on the extent of the sequence variation that may take place in a viroid population and which may result in the formation of a quasi-species. Moreover, 2186 new PLMVd variants are reported, making this study the most productive to date (by more than an order of magnitude) in terms of the reporting of novel viroid sequences.

**INTRODUCTION**

Viroids are plant-restricted infectious agents composed of a 245-401 nucleotide circular RNA genome (for a review see [1]). They are non-encapsidated and do not code for any protein. Their genomes possess sufficient information to take over the plant's transcriptional machinery and produce progeny that spread throughout the entire plant resulting in specific diseases [2]. They are divided into two families based on the presence or absence of a conserved central region (CCR) in their genome. The *Pospiviroidae* family is characterized by the presence of a CCR, and its members accumulate in the nucleus. Conversely, the *Avsunviroidae* family is characterized by the absence of CCR. Additionally, its members self-cleave via a *cis*-acting hammerhead motif during their replication cycle which takes place in the chloroplast.

It is still not known how viroids elicit pathogenesis. However, recent studies have focused on the implication of the RNA interference machinery. For example, it has been demonstrated that a hairpin derived from *Potato spindle tuber viroid* (PSTVd) alone can induce the symptoms associated with viroid infection when it is introduced into tomato plants [3]. Moreover, *Peach latent mosaic viroid* (PLMVd) variants inducing the peach calico disease, as well as the Y-satellite RNA of *Cucumber mosaic virus* (CMV), can induce symptoms following the interaction of viroid-siRNA with a specific host mRNA, thus silencing the targeted genes through an RNA-induced silencing complex (RISC) mediated degradation [2, 4]. In the case of PLMVd, the primary, rather than the secondary, structure mediates the symptoms observed during the peach calico disease through the binding of viroid small interfering RNAs with specific host mRNAs, resulting in the downregulation of the targeted RNA. Furthermore, data obtained with PTSVd have shown that the closing of one of its specific loops (i.e. loop 6), after a substitution of only 3 nucleotides, abolishes the systemic trafficking of this viroid in *Nicotiana benthiamana* [5, 6], suggesting a correlation between its sequence and/or its structure and the infection. Thus, accurate knowledge of a viroid's sequence, and of the sequence heterogeneity of mutant sequences generated following infection, is a key aspect in the elucidation of its pathogenesis.

Since their discovery in 1971 [7], a relatively large number of viroid sequences have been reported in public databases such as the Subviral one [8]. Currently, there are 2800 sequences, representing a total of 34 different species, demonstrating the intra-host genetic variation that exists in the viroid populations. *Avsunviroidae* viroids like PLMVd are replicated by a proofreading-deficient DNA-dependent RNA polymerase that is redirected to use RNA as template [9, 10]. As a consequence, viroid mutation rates are the highest (2.5 x $10^{-3}$ per site per replication cycle) reported to date for a biological entity [9]. More than 300 distinct sequences of PLMVd have been reported to date. These two facts led several investigators to claim that PLMVd, and more generally viroids, take the form of a cloud of related sequences in the multi-dimensional space of sequences that can be designated as a quasi-species [11]. In such a space, each point represents one particular sequence.

All PLMVd sequence variants reported to date have been cloned from total RNA isolated from a single tree (or a group of trees) using small-scale sequencing. The genetic variability of these sequences that may be found in only one host has not yet been addressed, and the conventional sequencing approach cannot take into account the large number of sequences that can be found in a single tree. The small size of viroids offers the advantage that a single high-throughput sequencing (HTS) using the 454 technology can determine their full-length sequences [12]. Moreover, the advent of this technology makes it possible to quantitatively list more of the viroid genomes found in a single tree than can Sanger sequencing. Clearly, HTS is a powerful tool with which to investigate a population of molecules [13]. In this study, a pipeline that permits a "snap-shot" of a viroid population 6 months post-infection is described. Use of this novel technique resulted in the demonstration that the PLMVd population is in fact composed of several clusters of variants, and that the sequences show significantly more heterogeneity than was previously believed. In all, 2186 novel sequences (including one new functional natural hammerhead sequence) were discovered, which, to our knowledge, is more than one order of magnitude greater than any previous study.

## RESULTS

**Deep-sequencing of circular PLMVd conformers**

In order to evaluate the genetic heterogeneity found in a viroid population, an infection experiment consisting of the slash inoculation of a single PLMVd variant into a single peach tree was performed. Six months after infection, the viroid's genomic circular conformer was isolated from the total RNA and then amplified before being completely sequenced. More specifically, a dimeric head-to-tail transcript of the PLMVd.282 variant (accession number: DQ680690), called the master sequence in this study and bearing one additional adenosine that was inserted between nucleotides 107 and 108, was used to infect a GF-305 indicator cultivar peach tree. This variant was identified in a previous sequencing study [14] and has been shown to be infectious when inoculated into the GF305 cultivar (data not shown). During the transcription of the DNA template containing the head-to-tail insert of the viroid, the hammerhead self-cleaving motifs do not cleave to completion, leaving an uncleaved dimeric RNA product which was used for the slash-inoculation. Six months post-infection, the circular conformers were purified from the total RNA of infected leaves using a denaturing polyacrylamide gel (Figure 1A). In parallel, Northern blot hybridization of RNA samples previously separated by native gel (i.e. without 8M urea) electrophoresis confirmed that solely the circular conformers were recovered (data not shown). Subsequently, RT-PCR amplification of the recovered circular RNA species, followed by deep-sequencing, was performed (Figure 1A). Since no sequence information can be generated from the regions bound by the primers, two pairs of primers complementary to both the P7 (positions 204 to 244) and P3 stem (positions 92 to 134) regions, which are believed to be highly conserved, were used for the RT-PCR amplification (Figure 1B). As a result, the two sets of primers specific for the (+) polarity strand that were used were sufficient to provide information on the entire viroid genome.

The sequencing step generated two libraries representing a total of 787 678 reads. The raw data from both libraries was then filtered as described in Materials and Methods (see also Table 1 for both the P7 and the P3 libraries). The filtering allowed the removal of reads that either missed a primer region, contained ambiguous positions (i.e. N), or were smaller than 249 nucleotides (nt) in length. An additional final step removed 116 unique reads that were less than 70% homologous to the PLMVd.282 master sequence (for a total of 227 reads) – all of them were rRNA - that were amplified by the P7 primers. From these filtering steps, 49.4% and 38.5% of the unfiltered reads of the P7 and P3 libraries, respectively, were kept for further analyses. This provided a total of 339 184 reads corresponding to 56 835 unique sequences (Table1). A detailed analysis of the number of sequences for each occurrence can be found in Figure S1. The number of occurrences here defines the number of times that a specific read (or nucleotide sequence) was found in a library.

**Refinement of the sequence libraries**

The libraries resulting from the filtering step are comprised of completed and defined reads. That said, they may also include reads containing mutations introduced during either the RT-PCR reactions or the deep-sequencing (also called technical mutations). In other words, mutations that do not reflect natural variation and that must be removed before any analysis can be performed. Previous reports have shown some limited flexibility in the composition of the catalytic core of the hammerhead self-cleaving motif (see the 13 boxed nucleotides for a strand of one polarity depicted in Figure 2A) without significantly affecting the cleavage ability [15-17]. Such variation notwithstanding, the extremely high conservation of the nucleotides composing the catalytic core was used here to eliminate reads with sequence variations that are likely due to sequencing artefacts. Since only replicative intermediates (i.e. circular conformers) were isolated, any mutation that impairs self-cleavage was considered to be the result of either the RT, the PCR or the deep-sequencing reactions. Mutations located in the core hammerhead nucleotides were searched for in both the P7 and P3 libraries. Variations in the composition of the hammerhead's catalytic core were detected in reads having 13 occurrences or less in the P3 library, while some mutations were found in reads having 6 occurrences or less in the P7 library. The discrepancy observed between the two libraries can be explained by the fact that, for pyrosequencing, the mutation rate increases with the distance from the 5' end of the sequencing primer [12]. In the present case, the P7 primers were located closer to the catalytic core than were the P3 primers (Figure 1B).

A set of mutated hammerhead self-cleaving sequences, derived from the reads found in both libraries, was analyzed using minimal constructions (Figure 2B and C). The cleavage efficiencies of the resulting radiolabelled *cis*-acting RNA transcripts were monitored by *in vitro* transcription reactions. The hammerhead motif of the master sequence was considered as the reference sequence. It exhibited 90% self-cleavage during a one-hour incubation. From the P7 library, three hammerhead motifs, all derived from reads having 6 occurrences each, were tested. One of these (P7_6_1630) was not able to cleave (percentage of cleavage <3%), while the other two exhibited significant levels of self-cleavage (> 50%). The mutation that hampered the cleavage was the substitution G12A (see the nomenclature in Figure 2A). This mutation has also been found in three entries of the Subviral database, PLMVd.082 (-), ASBVd.054 (+) and ASBVd.056 (+), and was denoted as a sequencing artefact by others [15, 16]. In the case of the P3 library, 4 hammerhead sequences were tested. All exhibited self-cleavage at a level greater than 50%, with the exception of the motif found in the P3_8_1113 read (having 8 occurrences) that exhibited a limited level of self-cleavage (i.e. 9%). Subsequently, the self-cleavage efficiency under which the reads were rejected was arbitrarily set to 15%. In other words, all reads having 8 occurrences or less were discarded and not considered for further analysis. One read in the P7 library having 10 occurrences and possessing no mutation in the core nucleotides, but with a deletion of 5 nucleotides that resulted in a smaller stem I, exhibited a cleavage efficiency that was decreased to 20% (data not shown). Since the cleavage efficiency was over 15%, the threshold was maintained at 8 occurrences.

The removal of the reads with less than 8 occurrences resulted in a reduction of the total number of reads by 23.5% to 259 486. This reduced the number of unique sequences by 96.2% to 2186 reads. It is important to note that about 75% of the sequences that were removed were only once, that is to say they represented single occurrence reads (see Figure S1). This threshold may appear stringent, but it is likely much better to remove a large number of reads than it is to consider those that bear mutations due to experimental artefacts. That said, it is also possible that mutations located outside of the core nucleotides of the hammerhead may impair self-cleavage.

When both libraries were compared, the sequences composing the P7 library appeared to be more heterogeneous. It is interesting to note that the average number of mutations, relative to the master sequence (PLMVd.282) used for the inoculation, was 6.3 mutations, and that it ranged from 3 to 21 mutations in the P7 library. The average number of mutations is lower in the P3 library (4.6 mutations), but the range (2 to 56 mutations) is broader (see Figure S2). In fact, in the P3 library a small number of sequences (84 sequences) contained between 37 and 56 mutations relative to the master sequence. If these sequences are excluded, all of the remaining sequences possess between 2 and 9 mutations relative to the master sequence. The higher average number of mutations observed in the P7 library could originate from the P3 region. Sequencing data from this region are missing in the P3 library because the primers used also anneal to this region during the cloning step. Inversely, it may also suggest that the region bound by the P7 primers (P7 stem) is more homogeneous.

**Natural hammerhead ribozymes with core nucleotide variations**

In the cleavage assays described above (Figure 2), 3 hammerhead ribozymes bearing variations in their core nucleotides escaped the threshold since their respective sequences had 8 or more occurrences (or frequencies). The other mutant ribozymes may be the result of technical mutations. Ribozyme P3_10_887 possesses the mutation G8A and still retains the ability to cleave as well as the ribozyme P3_9_978 which possesses the mutation U4C. These two mutations have previously been found in three natural viroids present in the Subviral database, (PLMVd.183(+), PLMVd.40(+) and CCHMVd.024(+)), leaving only P3_13_676 as a new natural hammerhead ribozyme with a core nucleotide variation that has maintained the ability to cleave. The variation in question is the insertion of an uridine between G12 and A13 (Figure 2A and C). This kind of modification to the core is expected to be easier to accommodate because the conserved GAAA region could "protrude" outside of the core, thereby resulting in a minimal structural rearrangements [17].

**Analysis of the collection of PLMVd sequence variants**

Taken together, the two libraries included 2186 novel PLMVd variants that were deposited in both the NCBI (accession numbers JX478274 to JX480459) and Subviral databases. Surprisingly, the master sequence used initially to inoculate the tree was not found in any of the libraries. The majority of the important mutations found in high occurrence reads were the results of substitutions, with a lesser number being deletions. The mutations were located all around the PLMVd genome. Merging of the sequences from both libraries revealed that 116 out of 254 positions (i.e. without considering the two regions recovered by the primers) were perfectly conserved. This corresponds to a 45.7% level of conservation (see Figure S3). Included in this number are 22 residues associated with the conserved sequence of the hammerhead motifs of both polarities, most of the P1 stem, one strand of the P5 stem (nucleotides 149-155) and most of the P10 stem including the GNRA (GUGA in this case) tetraloop that caps this stem. The latter was the only conserved loop of the viroid. In the case of the latter three groups, the contributions to the life cycle of the viroid of these highly conserved nucleotides are not known. Moreover, short stretches of conserved nucleotides, as compared to the mutations that were retrieved in a large number of the positions (i.e. 138 positions), were also observed. Another way to look at the sequence variation is in terms of the two distinct libraries. When this is done, the data from the P3 library shows that 169 positions out of 295 (excluding the region covered by the primers) were perfectly conserved, which corresponds to a 57.3% level of conservation. In the case of the sequences composing the P7 library, 203 positions out of 297 were fully conserved (i.e. corresponding to a 68.4% level of conservation). Finally, the most abundant reads in both the P7 and P3 libraries have 13 958 and 19 506 occurrences, respectively (see Figure S1). Together, these data clearly show that PLMVd is a mutation prone RNA genome.

**Clustering of PLMVd sequences**

In order to analyse the important number of reads, and to learn about the evolution of PLMVd after inoculation, a novel hierarchical (DHCS) algorithm [18] designed for the general purpose of clustering categorical sequences was employed. A unique advantage of DHCS is that it relies on an advanced statistical model, called Conditional Probability Distribution (CPD), to characterize a group of sequences, and to estimate similarity between an individual sequence and the group. This allows DHCS to by-pass sequence alignment for similarity computation, and makes it capable of dealing with various kinds of biological sequences, regardless of whether or not they are simple to align. As the first top-down divisive hierarchical algorithm for sequence clustering, DHCS improves several aspects of the clustering process. It bypasses comparisons of individual sequences in the estimation of pair-wise similarities during the clustering process, and yields statistically significant clusters. It offers an automatic method for estimating the number of clusters, and permits the generation of a hierarchy of clusters. In [18], it was shown that DHCS is effective for the clustering various sequence types including protein sequences.

Technically, DHCS owes its clustering capacity to a two-tier Markov model. The Conditional Probability Distribution model [19-21] used in DHCS to represent a group (cluster) of sequences is in fact a general Markov model of variable orders. It has the potential to explain the sequence generation process of a group by exploring significant motifs common to the group and to naturally integrate these motifs in a similarity estimation. That said, the application of the CPD to clustering presents computational difficulties in initializing clusters and in estimating the parameter values of the CPDs. The two-tier Markov model developed in DHCS overcomes these difficulties by using a first-order Markov model which yields a vector representation of each sequence in a group. The vector representation facilitates the splitting of the group as a solution to an optimization problem. The groups obtained from the optimized splitting are then used to initialize the CPDs, and are then optimized via a final refinement phase.

The two-tier Markov model is recursively used in DHCS to create the hierarchy of clusters. Starting with the entire set of sequences, DHCS makes successive bisections of the clusters, generating a binary tree. The nomenclature used is as follows: the name of each child node or cluster obtained after splitting is formed by adding 'O' for the left cluster and 'I' for the right cluster. For instance, "OOO" represents the left cluster from the bisection of cluster "OO" (Figure 3A). Each bisection (Figure 3B) is initialized by using the Fuzzy-Multiple Component Analysis (F-MCA) [22, 23] on the vector representation of the sequences produced by the first-order Markov model. The two groups are then optimized according to this first-order Markov representation, and the Chi-square distance between each sequence and each group is then used, producing two refined clusters. Finally, a variable-order CPD model for each of the two clusters is built depending on the motifs discovered from each cluster and on their statistical significance.

The application of the DHCS algorithm led to the grouping of the reads from the P7 library into 8 clusters, and to those from the P3 library into 7 clusters. An analysis by sequence comparison was then performed in order to compare each cluster. For each, the representative sequence (the sequence that is the center of the swarm, that is to say the other sequence possessing the minimum total dissimilarity to all of the sequences of the same cluster) was determined. Firstly, all of the sequences composing the cluster were multiple aligned in order to render the sequences the same length. Secondly, the resulting sequences (which for the shorter ones now have "-" at some positions) were then aligned pairwise. For each sequence, the mutational distances to other sequences of the cluster, in terms of the median number of mutations, were calculated. Finally, the sequence that possessed at the minimum number of mutations with respect to other sequences was retained as the representative of the cluster. This analysis revealed that all of the clusters, except one, possessed a median number of mutations relative to the representative of either 2 or 3. Some clusters did contain outliers possessing more than 12, but less than 17, mutations (Figure 4) as compared to the representative. Indeed, only the cluster IO from the P3 library exhibited a median number of mutations of 16.

Attempts were made to identify the mutations that distinguish each cluster, and to summarize the events that took place and lead to the observed differences in the fitness (explained below) of the clusters by performing a multiple alignment of the representatives of each cluster of P7 library (see Figure S4). This study, the biological fitness of a given sequence is approximated by the frequency with which it appears in the population, as seen through the optic of the sequenced data. Thus, high occurrence reads are considered fitter than low ones. The fraction of sequences that bear key mutations (i.e mutations that are found in the representative sequences of each cluster when they are aligned together) was compiled for each cluster and is presented in Table 2. Sequence proximity (i.e. the number of mutations difference from the master sequence) was used as a molecular clock. It is reasonable to imagine that different positions in the viroid sequence evolve with different mutation rates due to functional and structural constraints. However, in the present analysis, the evolutionary events were placed in order based on the distance in sequence space. The representative sequence of cluster P7-IOO was the closest to the master sequence and possessed 4 key mutations (Figure 5A). This cluster is of medium size, containing 81 different sequences for a total of 16 735 occurrences. Its representative sequence has an occurrence of 11 079 and possesses two deletions (ΔU290 and ΔC117) and two substitutions (C138U and C148A) that favour the closing of stem P4 (Figure 6). The first deletion was located at the cleavage/ligation site of the (+) polarity's hammerhead self-cleaving motif. The other deletion was in the deletion of C117, which is located in the upper part of P3 stem. According to previous mapping data on another variant (PLMVd.034) using different ribonucleases, this cytosine is bulged out from the structure [24].

Two clusters arose from the cluster discussed above, P7-IIO and P7-OII. The first was of low frequency with a representative sequence that accounts for a frequency of only 0.6% of this cluster. It is a smaller cluster limited to 90 different sequences (2164 occurrences) in which the representative sequence is characterized by the deletion of C118, which is located in the P3 stem. This mutation is a rare event (<10%) in the cluster, although the representative sequence possesses it. Most of the sequences (51%) include a C to U mutation at position 118 (i.e. C118U). One interesting mutation is G31A (described below) that is present in 33% of the sequences. Overall this cluster looks to be less fit than the previous one (i.e P7-IOO). However, the mutations G31A and C307U, when combined together, generate the fittest cluster (P7-OOO, see below). The next cluster, P7-OII, is characterized by the mutation C118U, and contains 166 different sequences for a total of 18 988 occurrences. It seems to be a pivotal cluster since 3 clusters arise from it. A first mutation, U296C, leads to a cluster of low fitness variants named P7-IOI. This small cluster contains only 52 different mutations and 1 035 occurrences. From this, the mutations G31A and C307U, both of which are located in the hammerhead sequences, lead to cluster P7-OOO. This is the largest cluster (and by definition the fittest), containing 316 different sequences and over 56 000 occurrences. The third cluster originating from the cluster P7-OII is characterized by the G104A mutation that is located in the upper part of the P3 stem. This mutation appears to hamper the fitness because its acquisition decreases the number of occurrences from 18 988 to 6 174. However, it seems that when the cluster acquires the two previous mutations (C307U and G31A), the variants become the fittest, as is illustrated by the fact that the number of occurrences increased from 6 174 to the 14 956 seen in cluster P7-OOI, clearly supporting the importance of these two mutations. Finally, the last cluster (denoted P7-III), originates from cluster P7-OOO and is characterized by the mutation A337G. This mutation, located in the loop of the P11 stem, seems to lower the fitness of the variants. It is also possible, however, that the cluster was just emerging since the RNA sample was taken after only 6 months of infection.

Similar analyses can be performed for the sequences of the P3 library (Figure 5B), as well as for the representative sequences for each cluster (see Figure S5 and Table 3). The sequences appear to be organized differently, most likely because genomic data for P3 stems that contains important mutations are not available. It is noteworthy that the mutations that lead to the fittest cluster (P3-OII) are the same ones observed in the P7 library. Clearly, both libraries demonstrated a good similarity. Like the P7 library, the mutations ΔU290, G31A, G307U, C138U and C148A were identified as key mutations in the P3 library (Figure 6). Interestingly, the P3 library included the largest cluster (P3-OII), in terms of the number of occurrences with 70 169 from 288 unique sequences, as well as the smallest (P3-IO) which contains only 4 different sequences that account for 84 occurrences (Figure 5B). The sequences exhibited between 37 and 56 mutations relative to the master sequence. This represents an almost 20% level of divergence in the same host, a level that has never been reported for a viroid. The mutations are scattered through out the entire sequence, with the exception of the region bound by the P7 primers that this is the most conserved region of the viroid and within which only one mutation was found (C236U; Figure S6). Interestingly, these 4 sequences, which are ~81-87% homologous to the master sequence, are not found in the P7 library even through the mutations would not have impaired the binding of the P7 reverse primer during the RT reactions.

Importantly, the existence of this cluster demonstrates that PLMVd is subject to important sequence heterogeneity.

In order to compare both libraries, the number of occurrences as a function of the position of the mutations was analyzed (see Figure S7). The same mutations were found to be present with virtually identical frequencies in both libraries.

**Mutations located in the vicinity of the hammerhead self-cleaving motifs**

Examination of the sequences forming the clusters with both the highest and the lowest numbers of sequences revealed the presence of a number of mutations located in the vicinity of the hammerhead motifs, that is to say within the P11 stem-loop. When the sequences participating in the formation of the hammerhead motifs of either the (+) or the (-) polarity were folded into their active secondary structures, a significantly high level of co-variation of the base-paired positions of each stem was observed (data not shown). This observation unambiguously demonstrated that these base-pairs are important, but that their identities are not. Moreover, it supports the notion that a strong selective pressure for the conservation of the hammerhead motif's secondary structure occurred in the infected plant. The analysis of the hammerhead sequences also revealed that many of the mutations are located in the loop regions, as well as in or near the catalytic core, although the latter are present in small amounts. It was tempting to speculate that the latter mutations may influence not only the self-cleavage activity, but also the quantity of accumulated copies of a given sequence in the plant. In order to investigate whether or not mutations acquired by the master sequence can modulate the fitness of the variants through self-cleavage, a cleavage experiment was designed that included three chosen hammerhead sequences: the master sequence (PLMVd.282), the representative of the fittest cluster (it is the same sequence in both libraries) and the representative sequence of the cluster P3-IO (Figure 7). The hammerhead self-cleavages from both the (+) and (-) polarity strands of these variants were tested via *in vitro* transcription (Figure 7). Investigation of the self-cleavage reaction during the transcription, rather than after, is more relevant because it better reflects the cellular process of viroid replication. The experiments were performed in triplicate, and figure 7 shows an example of a denaturing gel for the hammerhead sequences of both polarities. Interestingly, the hammerhead from the (-) strand of the representative of cluster P3-IO was the most efficient, with ~70% of self-cleavage after 1 h, while that of the (+) strand self-cleaved to only 10% under the same conditions. When the hammerheads of both the master sequence and the fittest cluster are compared, no significant differences were recorded (all four sequences self-cleaved at ~60-65% after 1 hour of incubation). Clearly, the level of self-cleavage alone cannot explain the occurrence of a given variant in the libraries, because if this was the case the best cleavage percentage would have been observed for the fittest variant and the lowest for the master sequence.

**DISCUSSION**

The aim of this work was to explore the genetic variability of the circular conformers of a quasi-species like PLMVd following the inoculation of a single sequence variant into a tree. The use of two primer pairs permitted the gathering of genetic data from the entire genome of the viroid six months after infection. The subsequent sequencing efforts were concentrated on the circular conformers in order to ensure that the information obtained came from active variants.

A novelty in the approach described here is the use of the extremely high conservation level found in the core nucleotides of the hammerhead to determine a threshold for the elimination of the mutations introduced by the RT, the PCR or the HTS stages. This is key to preventing the introduction of biases in the subsequent clustering step. The hypothesis used was that if a read contained a mutated hammerhead motif that prevented self-cleavage, it cannot be active in the cell and most likely was the result of experimental artefact. Since a hammerhead sequence that was unable to self-cleave efficiently (<15% of efficiency) was found in a read having up to 8 occurrences, this level was used as the threshold. Consequently, all sequences with 8 occurrences or less were removed from the analysis. This approach led to the discovery of a new natural hammerhead ribozyme with a core variation that retained the ability to self-cleave.

This arbitrary threshold definition does not take into account other aspects of the viroid replication cycle, and may be too stringent. For example, it is possible that a viroid possessing a relatively inefficient self-cleaving motif could be efficient in another step of the rolling circle replication (i.e. polymerisation or ligation), or that it may possess RNA motifs that could help it to be efficiently transported throughout plant by binding to specific proteins. A pragmatic choice of a threshold value was made since it is known that RT, PCR and HTS can introduce mutations. Importantly, the consideration of the highly conserved sequence of the hammerhead motif offers a reasonable means of determining a threshold.

Even using this stringent threshold, over two thousand new PLMVd sequences were retrieved in this experiment. This number is more than an order of magnitude larger than any previously reported sequencing effort for a viroid. The analysis of the libraries revealed that the master sequence inoculated into the plant was not found, even in the reads below the threshold, clearly demonstrating that evolution occurred at a fast rate even though the viroid was latent. A similar observation was recently made when a dimeric transcript of a PLMVd variant that induces peach calico was inoculated into GF305 seedlings. Although small-scale sequencing was used, and only a small portion of the viroid genome was analyzed, the master sequence was not recovered among the progeny [2]. The sequences obtained in the present study using the oligonucleotides complementary to the P7 region unambiguously demonstrated that the P3 region contained a number of mutations and cannot be considered as a highly conserved region for subsequent RT-PCR amplification and cloning. The P7 region appears to be more appropriate for this purpose, although some mutations were observed in the sequences of the P3 library. Ideally, a suitable approach permits the linearization of the circular genome and the addition of linkers at both extremities prior to RT-PCR amplification using oligonucleotides complementary to the external linker sequences. This types of strategy would also permit the elucidation of the entire sequence of a viroid variant as region of the viroid would be used to bind the primers.

The developed pipeline also includes a novel divisive hierarchical clustering (DHCS) algorithm that permitted model-based and alignment-free clustering of the sequences. Seven clusters were retrieved from the P3 and eight from the P7 libraries. The representative sequence of each cluster was identified. Except for one cluster found in the P3 library, all of the sequences forming a given cluster are 2 or 3 mutations apart distanced from the representative sequence (Figure 4), clearly illustrating that the DHCS algorithm was very effective. This effectiveness is also highlighted by the identification of key mutations that discriminate the clusters through multiple alignments of the representative sequences of each cluster. Most of the discriminating mutations are found in the majority of the sequences (>60%), confirming that these mutations are important. In addition, the clustering step permitted the identification of the mutations that lead to variations in the fitness of the clusters, as measured by the total number of occurrences of the sequences within a cluster. For example, the fittest cluster of P7 library contains more than 56 000 occurrences (cluster P7-OOO), while the fittest cluster of the P3 library contains about 70 000 occurrences (cluster P3-OII). In both the P3 and the P7 libraries, the most abundant sequence is the representative of the fittest cluster. In P7 library, the representative sequence of cluster P7-OOO has 13 958 occurrences, while that of cluster P3-OII in the P3 library has 19 506 occurrences. If the sequences defining the regions of the P3 and the P7 stems are removed, both sequences are identical. The most abundant sequences account for 11.6% of the data in the P7 and 14.0% in the P3 libraries, which corresponds to the distribution found in a quasi-species population in which the most abundant sequence accounts for 15% (the remaining 85% being a multitude of different variants) [25].

Interestingly, the same mutations appear to be key ones in both libraries. It would be very interesting to investigate whether or not the fit variants can still be maintained. After all, this study only took of snapshot of the progeny present 6 months after infection. Moreover, examination of whether or not the hammerhead self-cleaving motif efficiency may be responsible for the fitness indicated that self-cleavage catalytic efficiency during *in vitro* transcription did not correlate with the level of accumulated PLMVd variants (Figure 7). According to two studies in the model plant *Arabidopsis thaliana*, the replication of viroids is necessary, but not sufficient, for infectivity [26, 27], suggesting that the limiting step is probably the movement of the viroid within the plant. Clearly, this scenario is far more complicated than the single step of hammerhead self-cleavage, although the contribution of this RNA catalytic step is an important factor affecting selective pressure.

In summary, this study allowed the development of an original pipeline combining molecular biology and bioinformatics methods for the analysis of deep-sequencing data in order to study a population of viroids. Applying this procedure to PLMVd infected RNA samples using various viroid strains, or after various interval of time post-infection, should shed some light on the evolution of a viroid population within a plant. Moreover, this work reports the largest number of viroid sequences for any given viroid to date, as well as clearly demonstrating that viroids are prone to high mutation rates in a single plant. In the present case, the most distant variants found in the libraries possess a 20% variation level when compared to the master sequence, a result which was unexpected. This puts into question the assumption that 90% of an RNA should be conserved in order for it to be a quasi-species. Clearly, further work is required in order to be able to revisit the notion of the sequence variation of a viroid.

**MATERIALS AND METHODS**

**Plant infection**

The variant PLMVd.282 was dimerized and cloned as previously described [28]. The RNA was transcribed from a *SpeI*-digested plasmid using the TranscriptAidTM T7 High Yield Transcription Kit (Fermentas) and following the manufacturer's instructions in order to generate the (+) polarity strand. The RNA was then purified by phenol:chloroform extraction followed by ethanol precipitation. The resulting RNA pellets were washed in 70% ethanol and then dissolved in loading buffer (95% formamide, 10 mM EDTA, pH 8.0, 0.025% xylene cyanol and 0.025% bromophenol blue). The samples were then fractionated through 5% denaturing polyacrylamide gels (PAGE, 19:1 ratio of acrylamide to bisacrylamide) in buffer containing 45 mM Tris-borate, pH 7.5, 8 M urea and 2 mM EDTA. The dimeric RNA strand was extracted from the gel during a 16 h elution in an elution buffer containing 500 mM ammonium acetate, 10 mM EDTA and 0.1% SDS. After elution, the RNA was ethanol precipitated, washed with 70% ethanol, dried and dissolved in ultrapure water.

A healthy GF-305 peach tree obtained from a peach seedling was grown in greenhouse conditions prior to its slash inoculation with the dimeric RNA of PLMVd.282 dissolved in 50 mM $KH_2PO_4$.

**Purification of circular conformers from total RNA**

Total RNA was extracted from the leaves of the infected tree using Qiazol (Qiagen) according to the manufacturer's protocol. The total RNA (10 μg) then DNase treated at 37°C for 90 min in a reaction containing 40 mM Tris pH 8.0, 10 mM MgSO$_4$, 1 mM CaCl$_2$ and 3 U of RQ1 DNAse (Promega) in a final volume of 20 μl. The DNA free RNA then was purified by phenol-chloroform extraction followed by ethanol precipitation. The resulting RNA pellet was washed in 70% ethanol, dried and finally dissolved in a 1:1 mix of water and loading buffer (95% formamide, 10 mM EDTA, pH 8.0, 0.025% xylene cyanol and 0.025% bromophenol blue) prior to being fractionated through a 5% denaturing polyacrylamide gel (PAGE, 19:1 ratio of acrylamide to bisacrylamide) in buffer containing 45 mM Tris-borate, pH 7.5, 8 M urea, and 2 mM EDTA. Along with the total RNA, a radiolabelled *in vitro* self-ligated PLMVd.282 (+) transcript (338 nt long) was also run on the gel. The gel band corresponding to the migration distance of this transcript was excised from the total RNA lane of the gel and the RNA eluted for 16 h in elution buffer described above. Following elution, the RNA was ethanol precipitated, washed with 70% ethanol, dried and dissolved in ultrapure water.

**Cloning of PLMVd circular genomes for deep-sequencing**

The purified RNA was reverse transcribed in the presence of either 20 μM of the P3rev primer (5'-TGCAGTGCTCCGAATAGGGCAN-3'), or 20 μM of the P7rev primer (5'-GTTTCTACGGCGGTACCTGN-3'), using Superscript III reverse transcriptase (100 U, Invitrogen). The purified RNA, the primer and dNTPs (0.5 mM) were combined in water and incubated at 65°C for 5 min prior to being snap cooled on ice for 2 min. The reverse transcription (RT) reactions were performed in 50 mM Tris-HCl pH 8.3, 75 mM KCl, 3 mM MgCl$_2$, 5 mM DTT, RNaseOut (10 U, Invitrogen) and 5% DMSO at 50°C for 60 min.

After RT, the samples were PCR amplified using two primer pairs. The first pair, the P7F-P7 primer (5'-ccatctcatccctgcgtgtctccgactcag**AGACGCACTC**TGGATTACGACGTCTACCCGN-3') and the P7R-Long primer (5'-cctatcccctgtgtgccttggcagtctcagGTTTCTACGGCGGTACCTGN-3'), was used on the cDNA produced with the P7rev primer in the previous RT. The second pair, the P3F-P3 primer (5'-ccatctcatccctgcgtgtctccgactcag**AGCACTGTAG**GTTCCCGATAGAAAGGCTAAN-3') and the P3R-Long primer (5'-cctatcccctgtgtgccttggcagtctcagTGCAGTGCTCCGAATAGGGCAN-3'), was used on the cDNA produced with the P3rev primer. The small capitals nucleotides in the primers are the Titanium A and Titanium B specific sequences for the forward and reverse primers, respectively. The presence of these sequences are required by the 454-deep-sequencing facility. The bold uppercase nucleotides contain the sequence of the MIDBarcode, as both PCR products were multiplexed for the deep-sequencing reactions, while the uppercase nucleotides are specific to PLMVd. The PCR reactions were performed on the RT products in the presence of both sense and antisense primers (1 µM each) using Phusion DNA polymerase (2 U, New England BioLabs) as recommended by the manufacturer. The reactions were performed in Phusion HF buffer that contained 2.5 mM MgCl$_2$, 3% DMSO and 0.2 mM dNTPs in a final volume of 100 µl and involved the steps: 98°C: 30 sec; 30 cycles of (10 sec at 98°C, 20 sec at 55°C, 15 sec at 72°C) followed by a final 7 min elongation step at 72°C. The PCR reactions were then separated on 1% agarose gels, the gel bands corresponding to the products of ~350 bp were excised and the DNA purified using Spin-X centrifuge tube filters (Corning Incorporated). The purified DNA products were ethanol precipitated, washed and their pellets dissolved in ultrapure water. The samples were deep-sequenced at Genome Québec (Montréal, Canada) on a 454 GS-FLX Titanium pyrosequencing platform.

**Filtering of the raw data**

Using a series of Perl scripts, unfiltered sequence reads from both sets (P7 and P3) were trimmed of both the 5' and 3' primer sequences. Reads that did not contain the primers were discarded, as were those containing uncertainties (i.e. N). Reads smaller than 249 nt (normally, full length reads are about 295 nt long) and reads with lower than 70% identity to PLMVd were not kept for the following steps.

**Clustering**

In order to cluster sequences, DHCS relies on both a new two-tier Markov Model and an iterative divisive hierarchical process. It initializes each division of the clustering process by assuming that sequences follow the first-order Markov model, and by using Fuzzy Multiple Correspondence Analysis. An advanced variable-order Conditional Probability Distribution model was then built for each of the two groups. The composition of each group was optimized using the Chi-square distance between each sequence and each group in order to move a sequence from one group to the other. The group division was iterated n times depending on the desired numbers of clusters and depths. The algorithm was also equipped with automatic stopping criteria. The DHCS algorithm effectively associates a sequence with the most similar group (for further details, see Xiong et al. [18]).

The DHCS was run on both the P7 and the P3 libraries. Three parameters were integrated: the Markov model's order, the subsequence tree depth and the significance ratio which represented the minimal occurrence of a significant pattern (subsequence). The more the order and subsequence tree depth were increased, the greater the complexity. The order was set to five, the tree depth to seven and the significance ratio to one.

*In vitro* **self-cleavage assays**

DNA templates of the hammerhead sequences were prepared using a PCR-based strategy that included two oligodeoxynucleotides. The sense oligodeoxynucleotide contains the sequence of the T7 RNA polymerase promoter, while the antisense oligodeoxynucleotide contains both the inverse sequence of the hammerhead tested and the inverse sequence of the T7 RNA polymerase promoter at its 3' end (see Figure S8 for the sequences of all of the oligodeoxynucleotides used). For each self-cleaving sequence tested, 200 pmoles of both the sense and the antisense oligodeoxynucleotides were used in a filling reaction with *Pwo* DNA polymerase (Roche Diagnostics) containing 0.2 mM dNTPs, 20 mM Tris pH 8.8, 10 mM KCl, 10 mM (NH$_4$)SO$_4$, 0.1% Triton X-100 and 2 mM MgSO$_4$ in a final volume of 100 µl. The reactions were subjected to 11 cycles of 30 sec at 94°C, 30 sec at 50°C, and 30 sec at 72°C. The resulting PCR products were ethanol precipitated, washed with 70% ethanol and then dissolved in ultrapure water prior to being used as templates for *in vitro* transcription. *In vitro* self-cleavage of the hammerhead ribozymes were monitored during transcription by performing the reactions in the presence of 10 µCi of [α-$^{32}$P]GTP (3000 Ci/mmol; New England Nuclear), purified T7 RNA polymerase (5 µg), RNAseOut (10 U, Invitrogen), pyrophosphatase (0.005 U, Roche Diagnostics), and the PCR product (2 µM) in a buffer containing 80 mM HEPES-KOH pH 7.5, 24 mM MgCl$_2$, 2 mM spermidine, 40 mM DTT and 5 mM of each NTP in a final volume of 50 µL at 37°C for 60 min. Upon completion, the reaction mixtures were treated with DNase RQ1 (Promega) at 37°C for 20 min, and the RNA then purified by phenol:chloroform extraction and ethanol precipitation. The resulting pellets were washed in 70% ethanol and dissolved in loading buffer (95% formamide, 10 mM EDTA, pH 8.0). The samples were then fractionated through 15% denaturing polyacrylamide gels (PAGE, 19:1 ratio of acrylamide to bisacrylamide) in buffer containing 45 mM Tris-borate, pH 7.5, 8 M urea, and 2 mM EDTA. The reaction products were visualized by exposure to phosphor imaging screens and revealed on a Typhoon scanner (GE Healthcare).

In the case of the time course experiments, the transcription reactions were carried out at 25 °C and 3 µl aliquots were withdrawn (at 1, 5, 10, 30 and 60 min) and immediately mixed with 15 µl of loading buffer in order to stop the reaction. Without further treatment, the samples were electrophoresed on a 15% denaturing polyacrylamide gel and exposed as described above.

**ACKNOWLEDGEMENTS**

Special thanks to Audrey Dubé and Olivier Parisi for technical assistance, in particular with the inoculations, to Dominique Lévesque for the generation of the plasmid pPD2-4 and to Tengke Xiong for his assistance with the DHCS algorithm and programs.

**FIGURE LEGENDS**

**Figure 1: Experimental scheme.** (A) A plasmid containing a dimeric head-to-tail arrangement of the sequence of PLMVd.282 was digested with *SpeI* and then used in an *in vitro* transcription reaction (step 1). The resulting dimeric transcript was isolated by excision following migration on a denaturing gel (step 2) (see dotted line on the gel). This dimeric viroid RNA was slash-inoculated into one GF-305 peach tree (step 3). Six months post-infection, total RNA was isolated from leaves (step 4) and fractionated on a 5% denaturing gel (step 5). Radiolabelled 338 nt self-ligated monomeric circular and linear monomer PLMVd RNAs were also electrophoresed as markers on the gel. The circular conformers were isolated from the infected cell RNA by excision of the gel band (dotted line) corresponding to the radiolabelled circular PLMVd from the gel (step 6). RT-PCR amplification was then performed using two primer pairs (P7 and P3) (step 7) prior to deep-sequencing of the PCR products (step 8). (B) Both forward and reverse P3 primers bind to the region from nucleotides 92 to 134, while the P7 primers bind to the region from nucleotides 204 to 244. The four primers are represented by dotted lines here. Each stem (P1 to P11) is indicated, as is the hammerhead self-cleavage site of the (+) polarity strand (arrow). The black and open boxes below and above the P11 stem identify the regions that compose the hammerhead core of the (+) and (-) polarity strands, respectively.

**Figure 2: *In vitro* cleavage assay of minimal versions of hammerhead sequences.** (A) Schematic representation of the consensus sequence of the catalytic core of the hammerhead sequences. The core nucleotides, identified by the boxes, are flanked by the helical stems I-III. The numbering is according to Hertel et al. [29]. (B) The cleavage assays were monitored during 60 min transcription reactions at 37 $^{o}$C, and were analyzed on a 15% denaturing gel. The non-cleaved full-length transcripts are indicated by the letters NC, while the cleaved products are identified by the letter P. At the bottom of the gel, a histogram displaying the cleavage percentage for each reaction is shown. (C) Schematic representations of the hammerhead motifs tested in B with their mutations identified. The different hammerhead ribozymes tested are identified by three terms as Library_Occurrences_Arbitrary number (for example: P7_6_1516).

**Figure 3: Generation of clusters by the DHCS algorithm.** (A) All of the sequences in each library were clustered using the DHCS algorithm developed by Xiong et al, [18]. (B) During the clustering, the sequences are grouped according to their relative similarities and each sequence is tested with the center of each cluster (I or O). It is still possible at this point that a specific sequence may be retrieved in another cluster. This step is repeated until the clusters become stable and the distance between each sequence and the center is minimal.

**Figure 4: Dispersion of the variants relative to the representative sequence of each cluster.** Box-and-whisker diagrams showing the dispersion of the variants relative to the representative sequence in each cluster of the P7 (A) and P3 (B) libraries. For each cluster, the blue box contains 50% of all of the sequences in the cluster. The red line shows the median of the number of mutations, and the little red crosses represent outliers of the cluster.

**Figure 5: Clusters and discriminating mutations.** (A) The clusters of the P7 library are represented by boxes containing the cluster's characteristics: the name of the cluster, the number of different sequences (Nb seq), the total number of occurrences (Total Occ) and the identification of the representative (Rep). The latter is described by three numbers separated by a ";" symbol. The first number refers to an arbitrary number. The second refers to the number of occurrences of the sequence and the third to the length of the sequence. In brackets, the fraction of the representative sequence in the cluster is shown. Each cluster is linked to the next by the most significant mutation(s) found following a multiple alignment of the representative sequences. (B) Same as in A, but for the P3 library.

**Figure 6: Key mutations detected in the representative sequence of each cluster**. The key mutations that characterize the different clusters are depicted on the accepted secondary structure of the PLMVd (+) polarity strand. Mutations specifically identified in P7 library are in red, while those associated with P3 library are in blue. The purple mutations were found in both libraries. The black arrows indicate the hammerhead self-cleavage sites for the strands of both polarities. The black and open boxes below and above the P11 stem identify the regions that compose the core hammerhead nucleotides of the (+) and (-) polarity strands, respectively.

**Figure 7: Cleavage efficiencies of the evolved self-cleaving hammerhead motifs**. (A) A typical denaturing gel showing the cleavage of the (-) strand hammerhead ribozymes of the master sequence (PLMVd.282), the representative of the fittest cluster (FIT) and the representative of the cluster P3-IO found in P3 library is shown. In each case, aliquots were taken at 1, 5, 10, 30 and 60 minutes during the transcription of their respective DNA templates in the presence of [$^{32}$P]UTP at 25$^{o}$C. On the right side, the mutations relative to the master sequence are depicted for each ribozyme. (B) Same as in (A), but for the (+) strand ribozymes.

**SUPPORTING INFORMATION**

**Figure S1: Analysis of the number of sequences of each occurrence found in both the P3 and P7 libraries.** This detailed analysis was performed on filtered reads.

**Figure S2: Number of mutations relative to the master sequence.** For each library, all of the sequences over the threshold (>9 occurrences) were analyzed in terms of their numbers of mutations relative to the master sequence.

**Figure S3: Conservation.** Secondary structure of PLMVd showing, in orange, the nucleotides that are 100% conserved when the data from both libraries are merged. The shaded nucleotides represent the regions bound by the primers.

**Figure S4: An alignment of the master sequence with the representative sequences of each cluster of the P7 library.** Each key mutation is identified by a box.

**Figure S5: An alignment of the master sequence with the representative sequences of each cluster of the P3 library**. Each key mutation is identified by a box.

**Figure S6: An alignment of the master sequence with the 4 sequences contained in the cluster P3-IO.**

**Figure S7: Types of mutations.** Graphs showing the number of occurrences of the different types of mutations, as well as their respective positions on the viroid's genome for both libraries. The grey boxes cover the regions bound by the primers from which no genetic data is available.

**Figure S8: Oligonucleotides.** DNA oligonucleotides used as templates for the *in vitro* self-cleavage assays. The italic nucleotides (in 3') represent the binding site for the T7 RNA polymerase promoter.

**Table 1 :** Analysis of the reads for both the P7 and P3 libraries obtained following deep-sequencing of the circular conformers RT-PCR amplified from the total RNA of the infected tree.

| Library | # of unfiltered reads | After filtering | | After threshold removal | |
|---|---|---|---|---|---|
| | | Total # of reads | # of unique reads | Total # of reads | # of unique reads |
| **P7** | 329 094 | 162 603 | 30 908 | 119 843 | 1 125 |
| **P3** | 458 584 | 176 581 | 25 927 | 139 643 | 1 061 |
| **Total** | **787 678** | **339 184** | **56 835** | **259 486** | **2 186** |

**Table 2 :** Fraction (percentage) of reads possessing the key mutations in each cluster of the P7 library.

|         | IOO  | IIO  | OII  | IOI  | OOO  | OIO  | III  | OOI  |
|---------|------|------|------|------|------|------|------|------|
| **ΔU290**  | 95   | 86   | 72   | 92   | 82   | 96   | 70   | 86   |
| **ΔC117**  | 99   | 99   | 94   | 99   | 99   | 99   | 99   | 99   |
| **C138U**  | 99   | 99   | 99   | 99   | 99   | 99   | 99   | 99   |
| **C148A**  | 99   | 99   | 99   | 99   | 99   | 99   | 99   | 99   |
| **C118U**  | 5    | 64   | 51   | 40   | 75   | 89   | 81   | 96   |
| **ΔC118**  | <10  | <10  | <10  | <10  | <10  | <10  | <10  | <10  |
| **U296C**  | <10  | <10  | <10  | 20   | <10  | <10  | <10  | <10  |
| **G31A**   | <10  | 33   | <10  | 19   | 99   | 23   | 99   | 99   |
| **C307U**  | <10  | <10  | 31   | 37   | 67   | 28   | 65   | 63   |
| **A337G**  | 20   | 23   | <10  | 30   | <10  | <10  | 56   | <10  |
| **G104A**  | <10  | 37   | <10  | <10  | <10  | 71   | 14   | 88   |

**The shaded boxes represent the mutations that give birth to the cluster.**

**Table 3 :** Fraction (percentage) of reads possessing the key mutations in each cluster of P3 library.

|         | OIO | OOI | OII | III | IIO | OOO |
|---------|-----|-----|-----|-----|-----|-----|
| **G31A**    | 56  | 72  | 91  | 60  | 99  | 99  |
| **C138U**   | 99  | 99  | 99  | 99  | 99  | 99  |
| **C148A**   | 99  | 99  | 99  | 99  | 99  | 99  |
| **INSU331** | <10 | 19  | <10 | <10 | <10 | <10 |
| **C307U**   | <10 | 36  | 99  | 22  | 99  | 99  |
| **ΔU290**   | 51  | 35  | 54  | 55  | 49  | 36  |
| **G37A**    | <10 | <10 | <10 | 33  | 15  | <10 |
| **C274U**   | <10 | <10 | <10 | <10 | 32  | <10 |

**The shaded boxes represent the mutations that give birth to the cluster.**

Figure1
Click here to download high resolution image

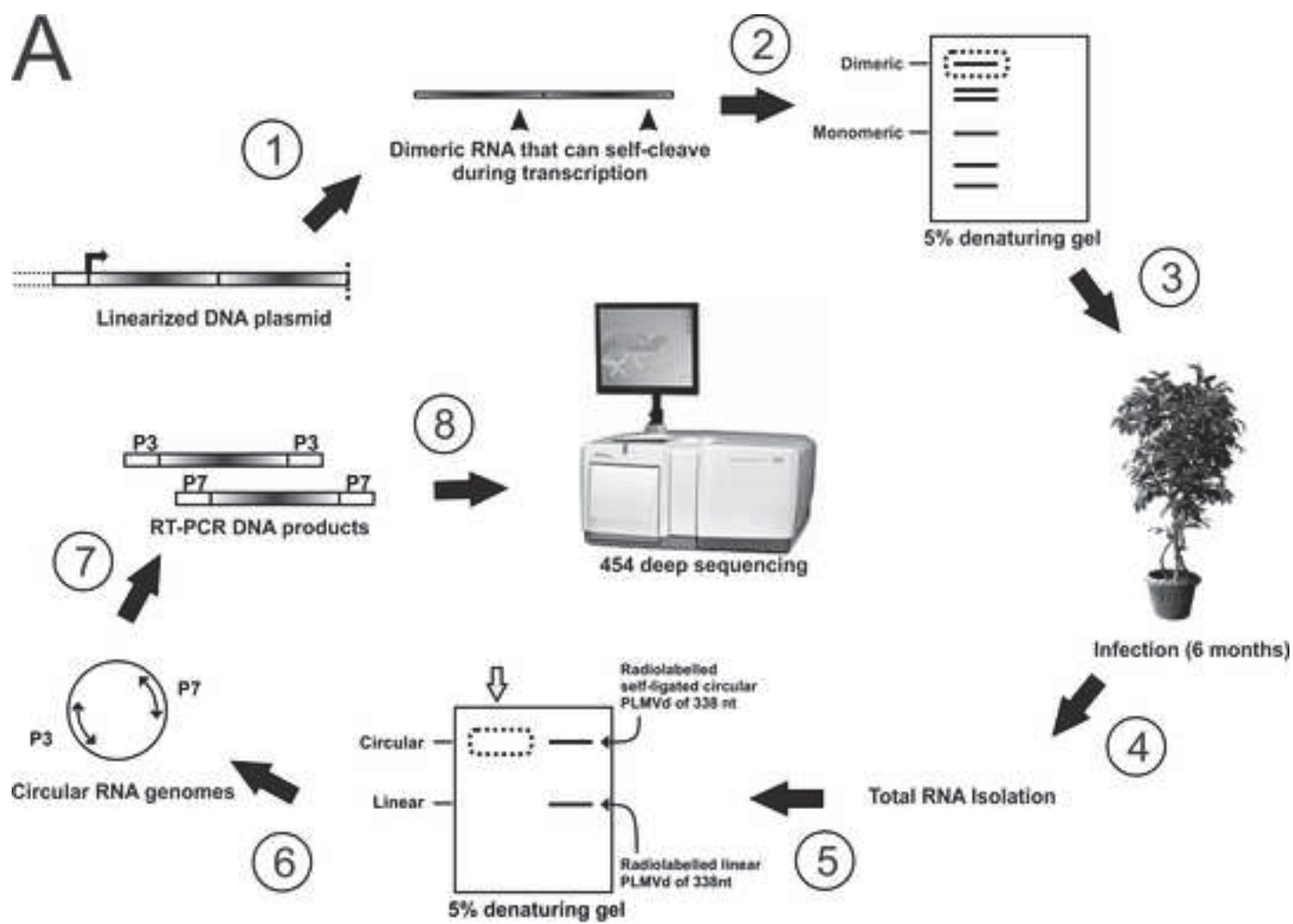

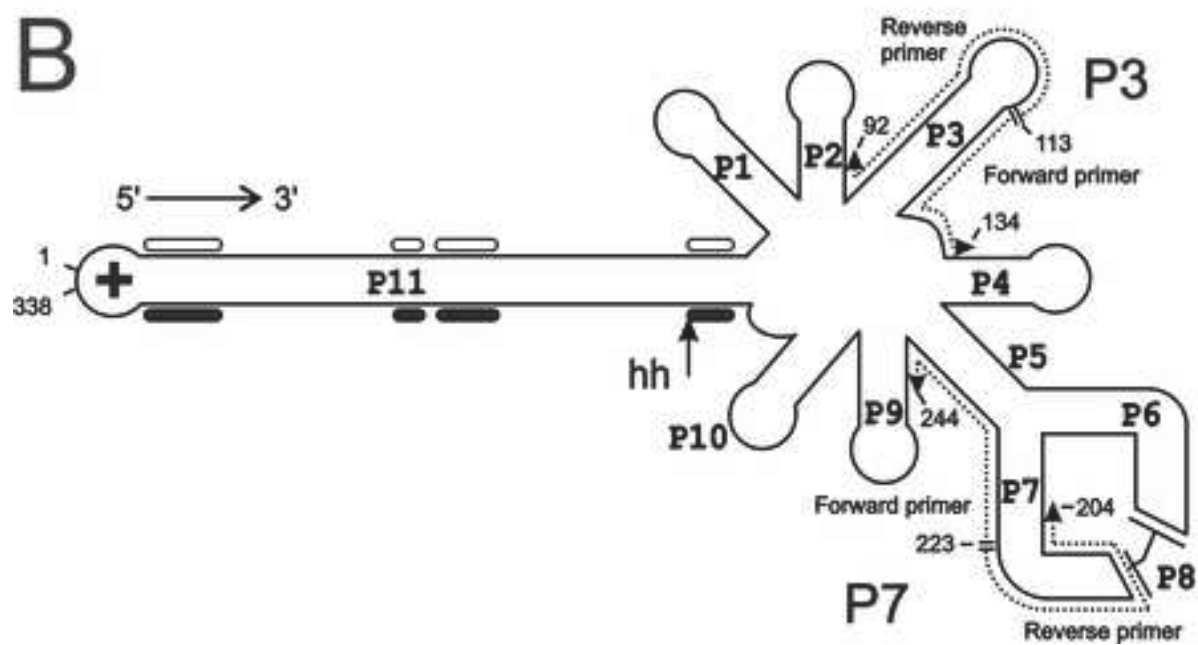

Figure2
Click here to download high resolution image

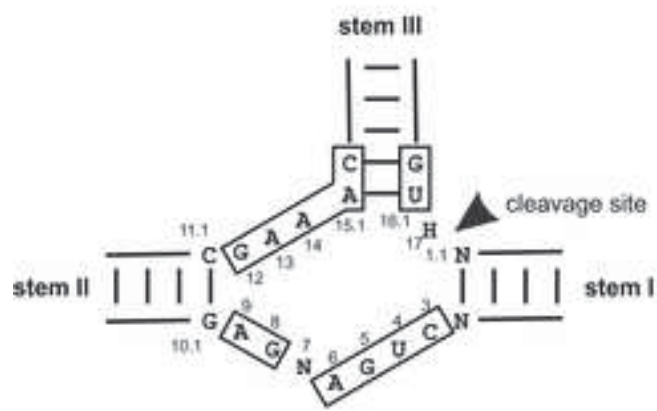
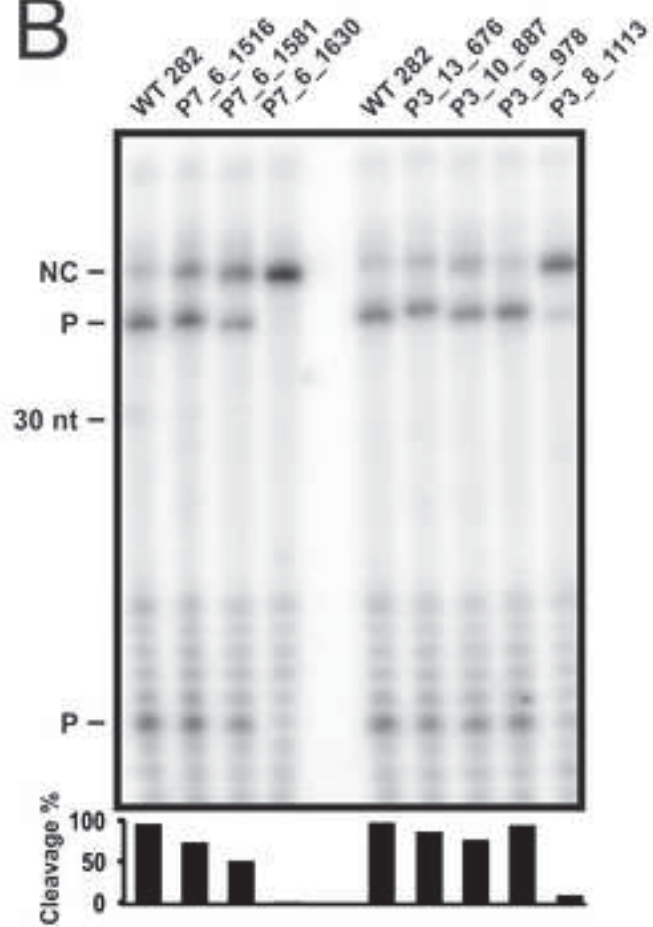
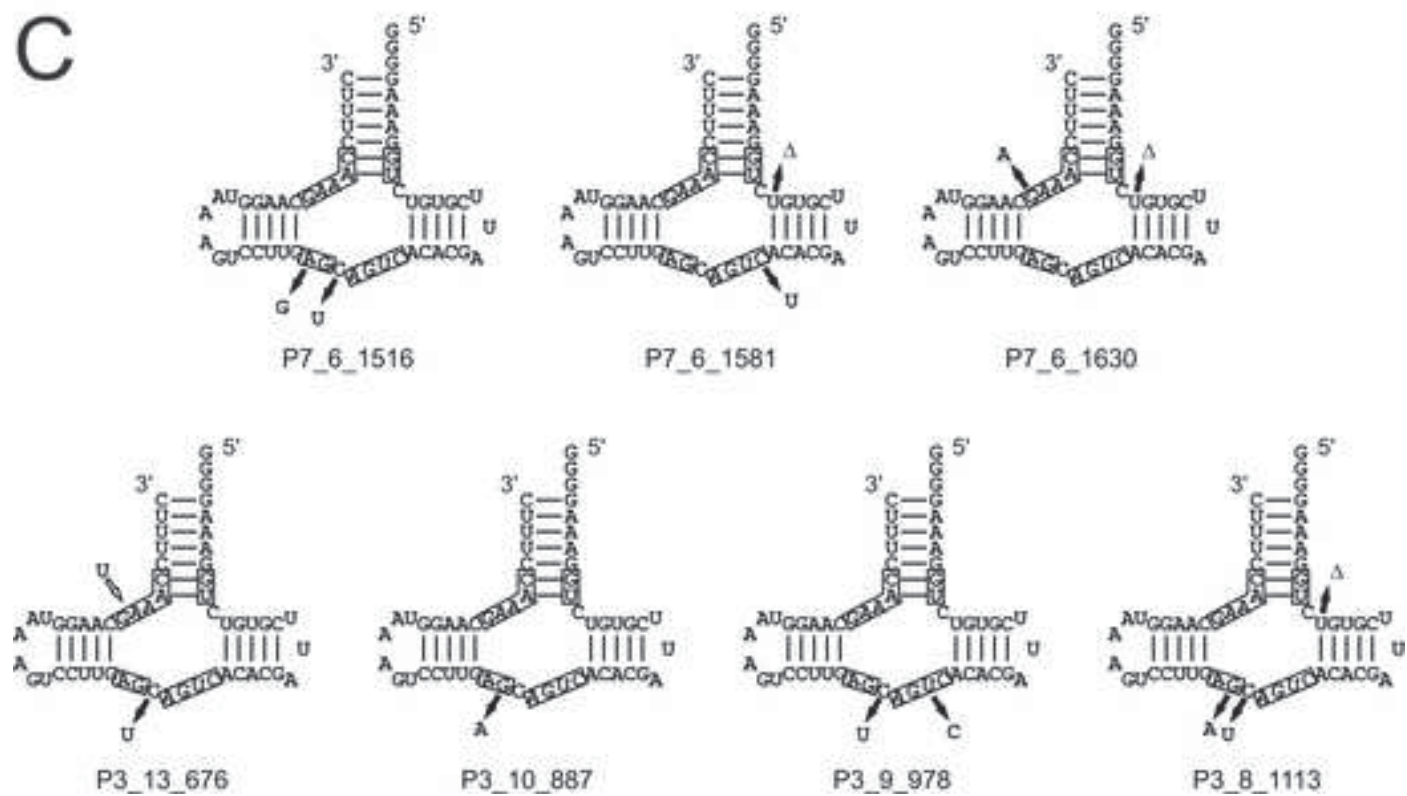

Figure3
Click here to download high resolution image

A
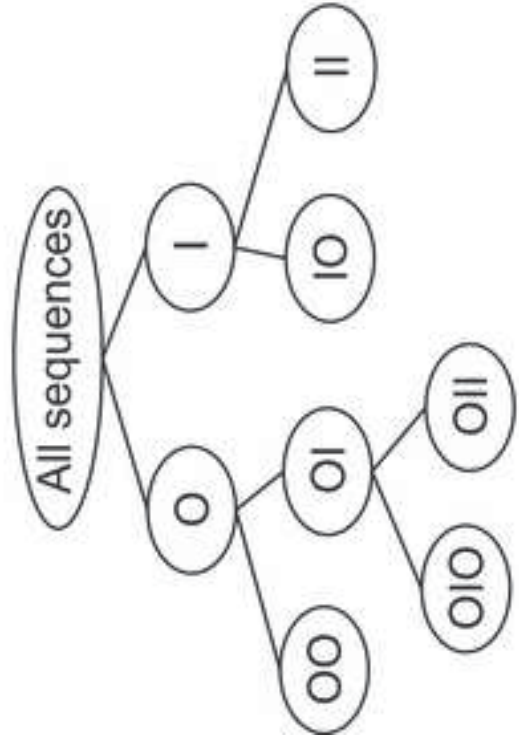

B
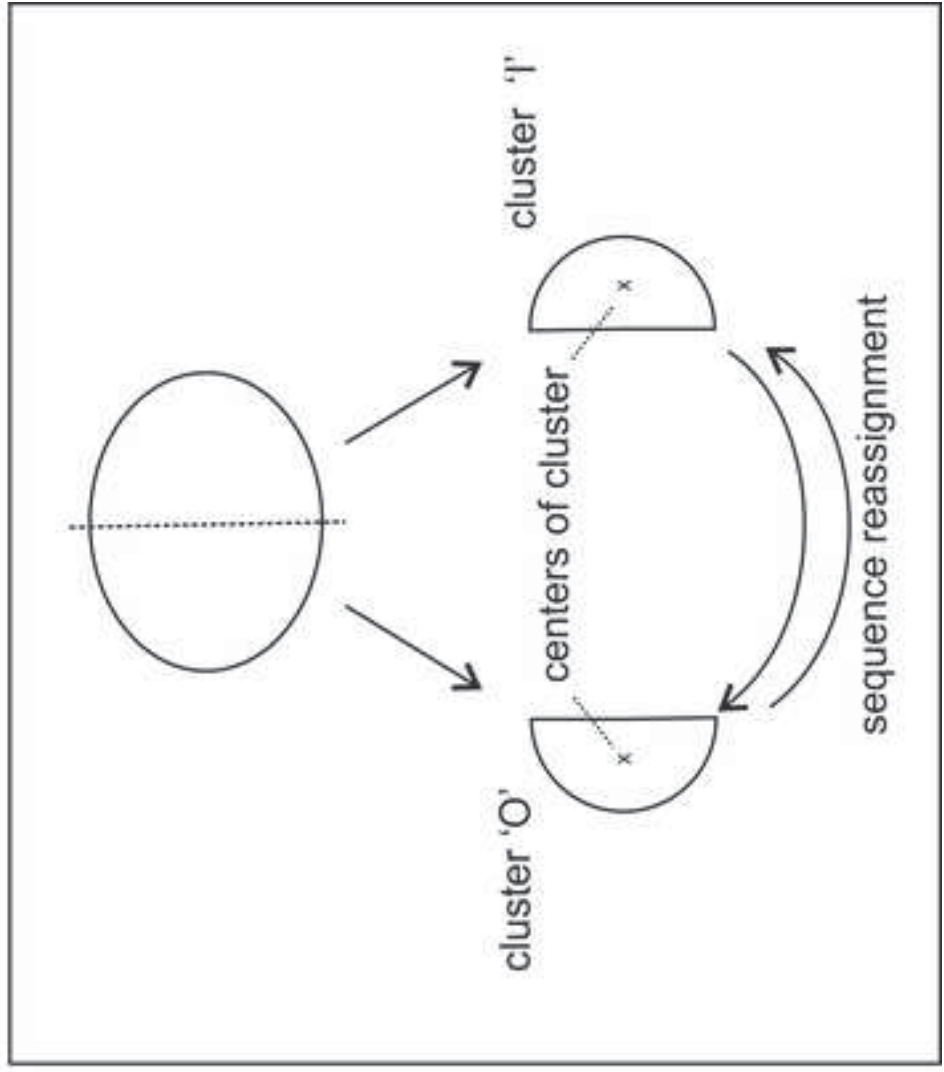

Figure4
Click here to download high resolution image

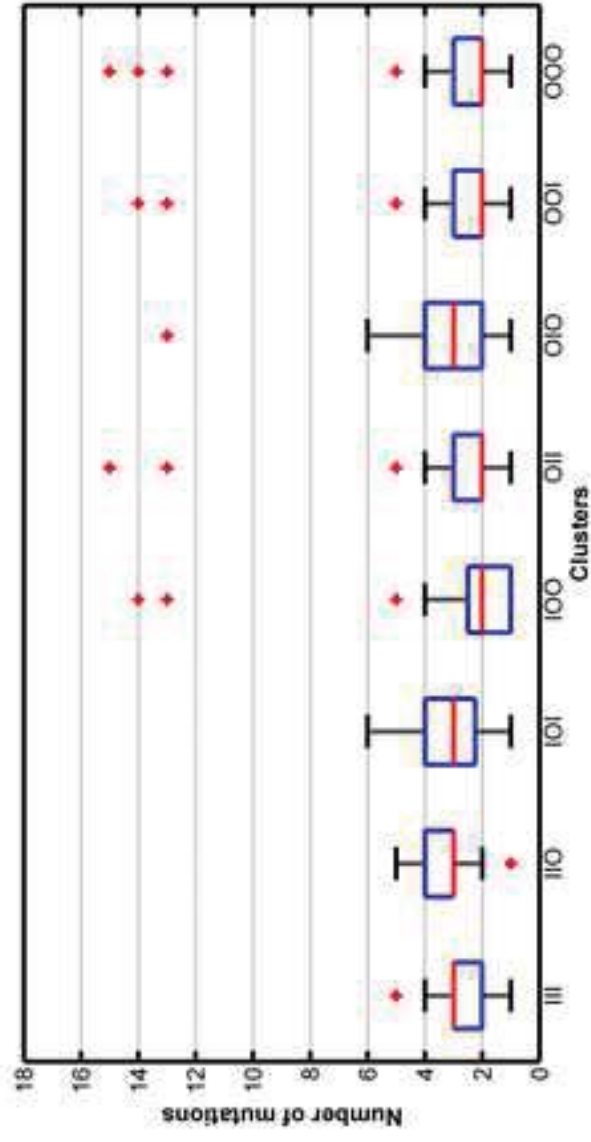
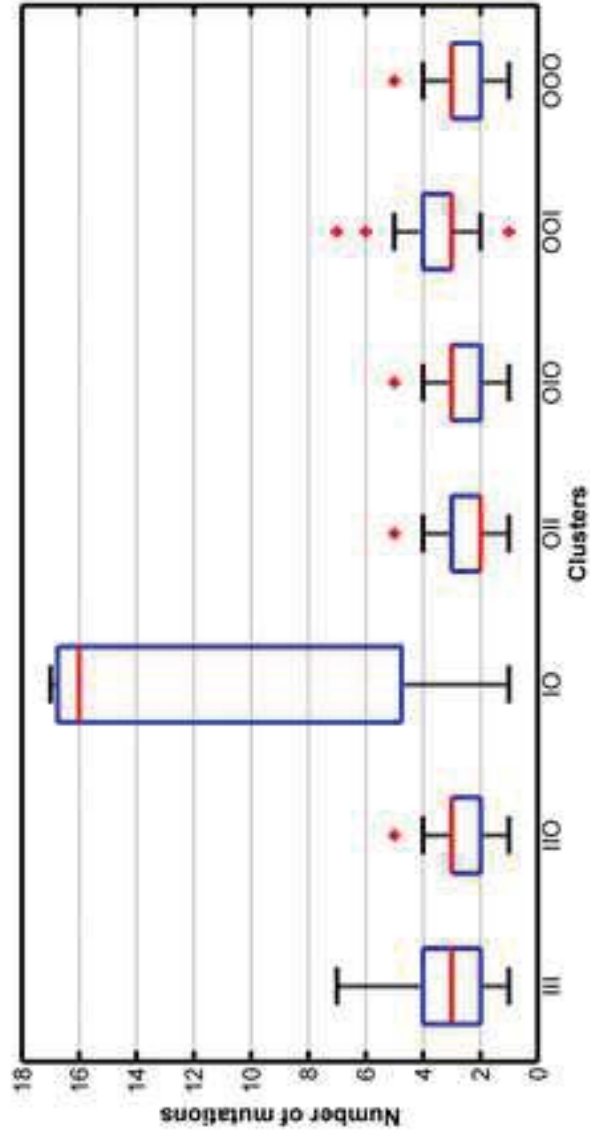

Figure5
Click here to download high resolution image

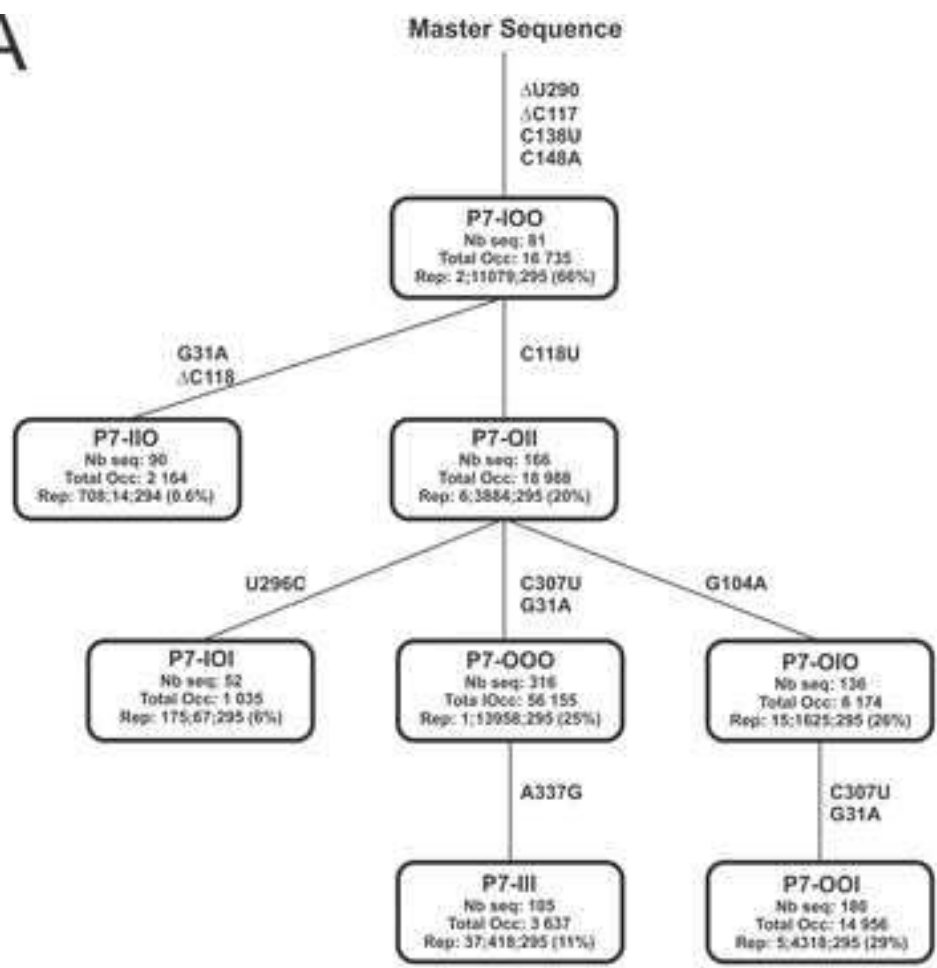
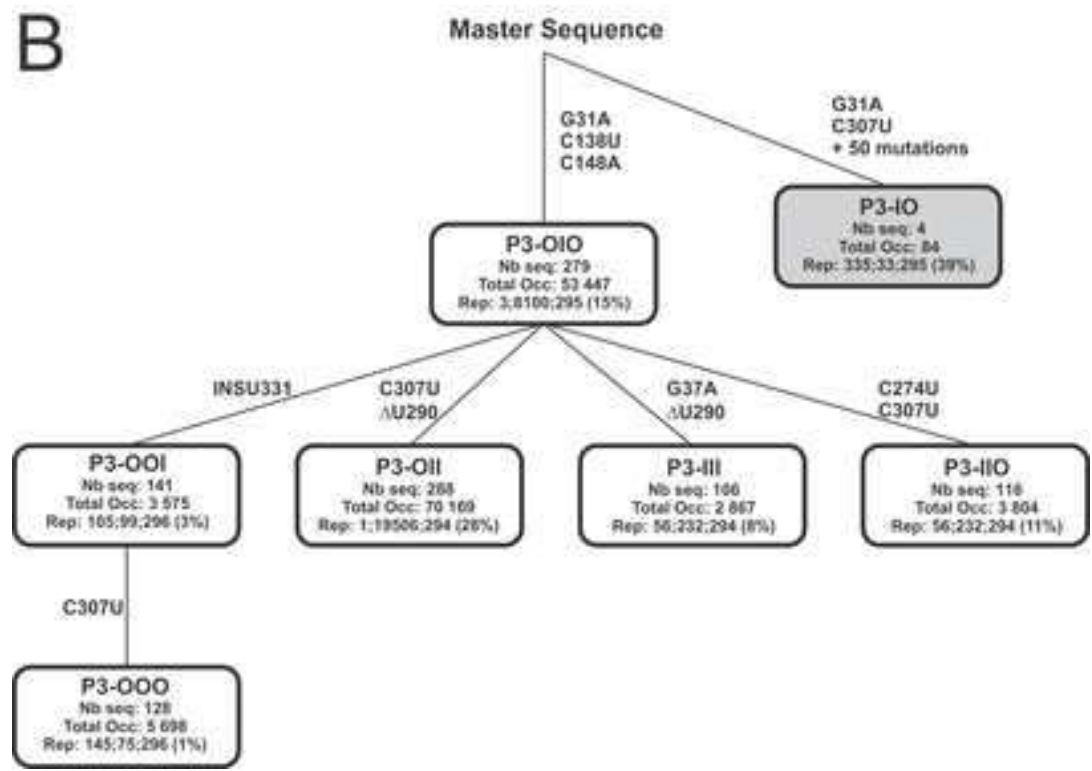

Figure6
Click here to download high resolution image

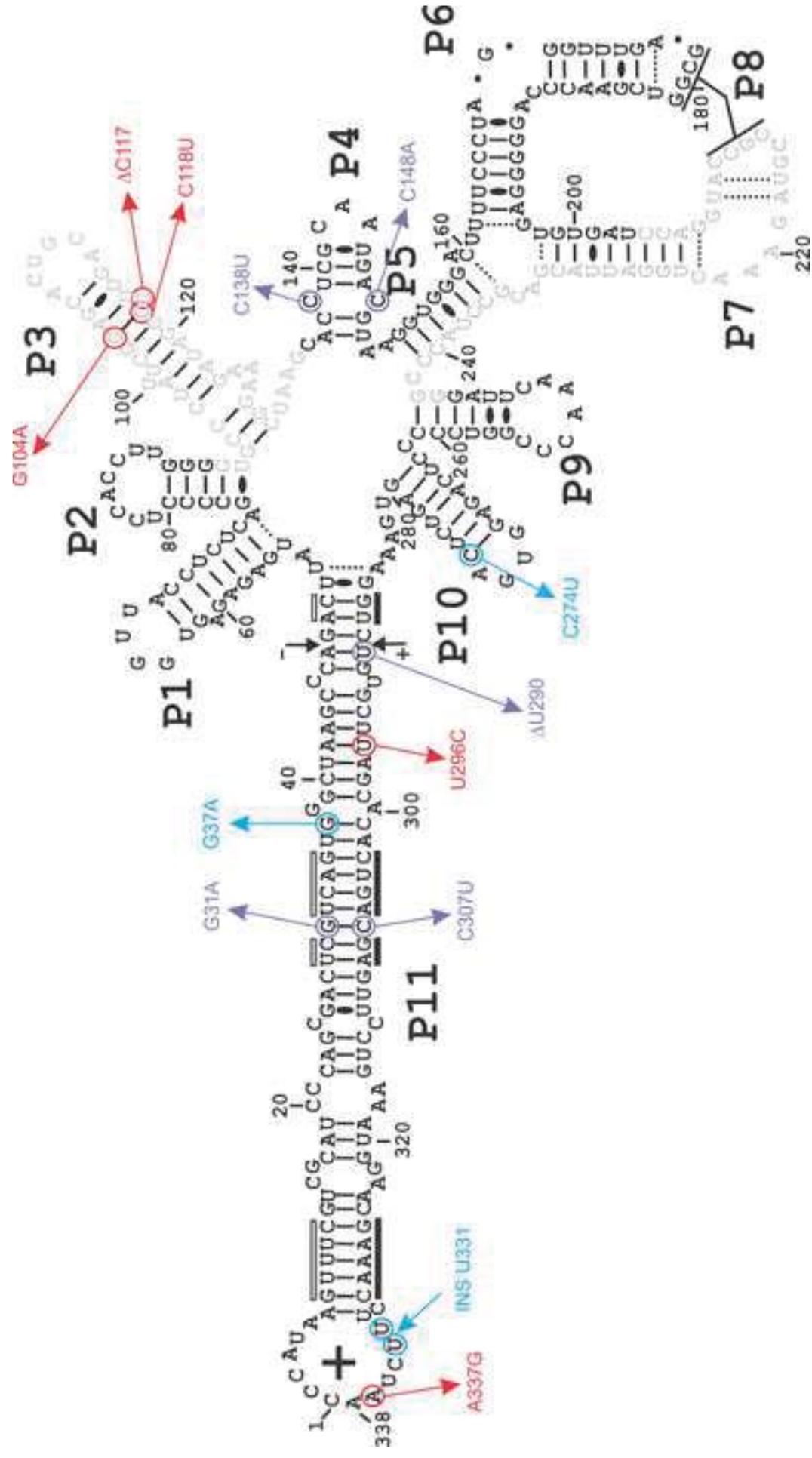

Figure7
Click here to download high resolution image

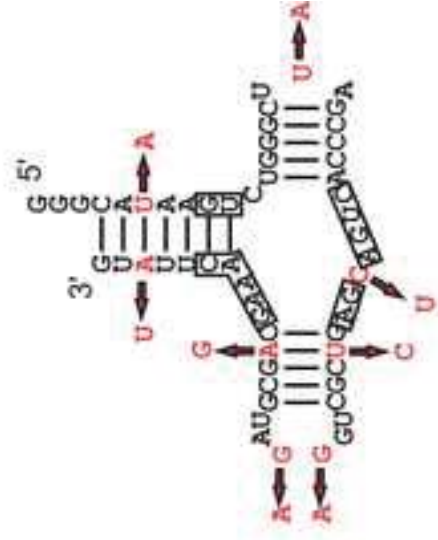
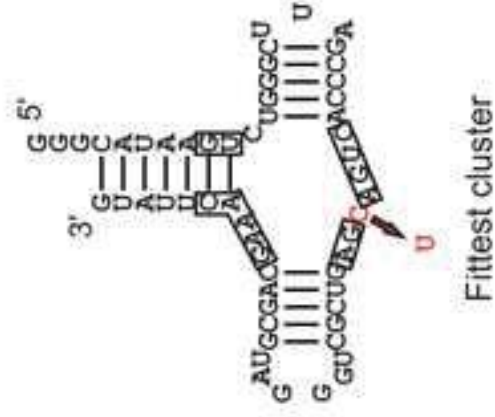
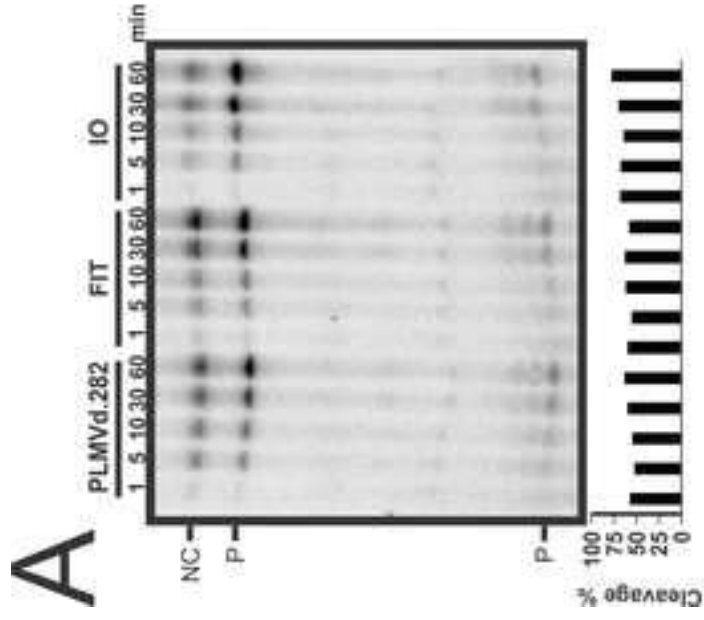

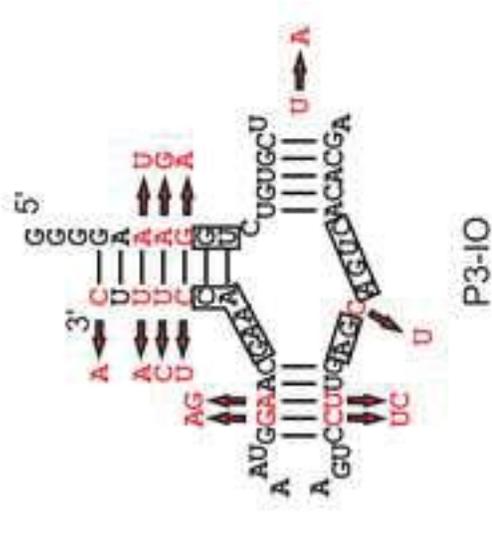
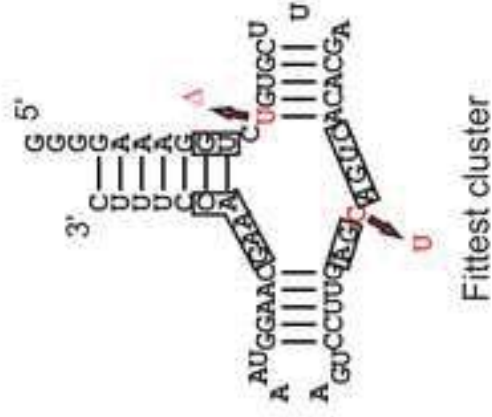
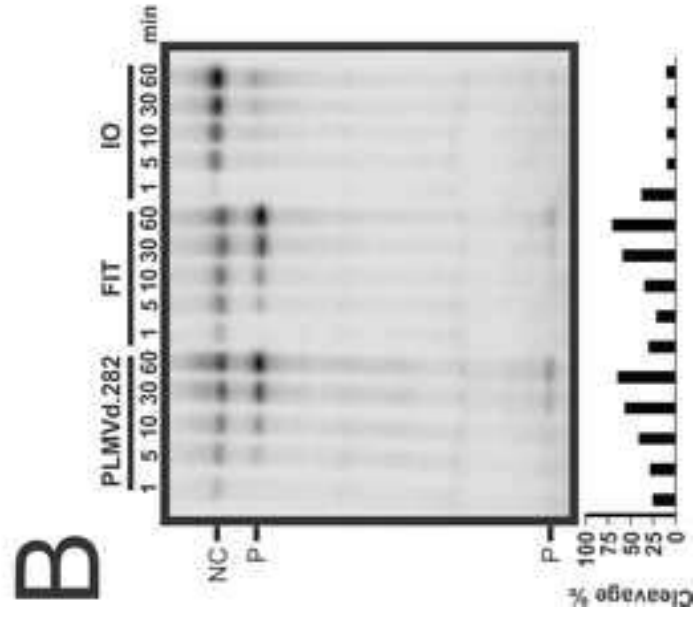

Supporting Information - Compressed/ZIP File Archive

Click here to download Supporting Information - Compressed/ZIP File Archive: Sup_Info.zip